\documentclass[11pt]{article}
\usepackage{amsmath,amsfonts,amssymb}
\usepackage[usenames]{color}
\usepackage{pstricks}
\numberwithin{equation}{section}

\newcommand{\nc}{\newcommand}
\nc\disp{\displaystyle}
\nc{\fh}{\hat{f}}
\nc{\muh}{\hat{\mu}}
\nc{\nuh}{\hat{\nu}}
\nc{\spos}[2]{\makebox(0,0)[#1]{$\sm{#2}$}}
\nc{\sm}[1]{{\scriptstyle #1}}
\nc{\bib}{\bibitem}
\nc{\al}{\alpha}
\nc{\g}{\gamma}
\nc{\G}{\Gamma}
\nc{\D}{\Delta}
\nc{\eps}{\epsilon}
\nc{\la}{\lambda}
\nc{\La}{\Lambda}
\nc{\var}{\varphi}
\nc{\pa}{\partial}
\nc{\nn}{\nonumber \\ }
\nc{\hf}{\frac{1}{2}}
\nc{\dz}{\frac{dz}{2\pi i}}
\nc{\bin}[2]{\left(\!\!\!\begin{array}{c} {#1}\\ {#2} \end{array}\!\!\!\right)}
\nc{\be}{\begin{equation}}
\nc{\ee}{\end{equation}}
\nc{\bea}{\begin{eqnarray}}
\nc{\eea}{\end{eqnarray}}
\nc{\bra}[1]{\langle {#1}|}
\nc{\ket}[1]{|{#1}\rangle}
\nc{\ketw}[1]{({#1})^{\phantom{a}}_{{\cal W}}}
\nc{\chit}{\raisebox{0.25ex}{$\chi$}}
\nc{\chih}{\raisebox{0.25ex}{$\hat\chi$}}
\nc{\Db}{\mbox{\boldmath $D$}}
\nc{\Hb}{\mbox{\boldmath $H$}}
\nc{\calH}{{\cal H}}
\nc{\calR}{{\cal R}}
\nc{\calL}{{\cal L}}
\nc{\calV}{{\cal V}}
\nc{\Hc}{{\cal H}}
\nc{\Rc}{{\cal R}}
\nc{\Lc}{{\cal L}}
\nc{\Vc}{{\cal V}}
\nc{\Ib}{\mbox{\boldmath $I$}}
\nc{\qb}{\bar{q}}
\nc{\vh}{\hat{v}}
\nc{\wh}{\hat{w}}
\nc{\Jh}{\hat{J}}
\nc{\Qh}{\hat{Q}}
\nc{\Ac}{\mathcal{A}}
\nc{\Bc}{\mathcal{B}}
\nc{\Cc}{\mathcal{C}}
\nc{\Dc}{\mathcal{D}}
\nc{\Ec}{\mathcal{E}}
\nc{\Ic}{\mathcal{I}}
\nc{\Jc}{\mathcal{J}}
\nc{\Nc}{\mathcal{N}}
\nc{\Oc}{\mathcal{O}}
\nc{\Sc}{\mathcal{S}}
\nc{\Xc}{\mathcal{X}}
\nc{\Yc}{\mathcal{Y}}
\nc{\Zc}{\mathcal{Z}}
\nc{\fus}{\mbox{}\,\hat\otimes\,\mbox{}}
\nc{\Sh}{\hat{S}}
\nc{\Sct}{\tilde{\mathcal{S}}}
\def\vvdots{\mathinner{\mkern1mu\raise1pt\vbox{\kern7pt\hbox{.}}\mkern2mu
  \raise4pt\hbox{.}\mkern2mu\raise7pt\hbox{.}\mkern1mu}}
\nc{\gauss}[2]{\left[\!\!\begin{array}{c} {#1}\\ {#2} \end{array}\!\!\right]}
\nc{\sbin}[2]{\left\{\!\!\!\begin{array}{c} {#1}\\ {#2} 
\end{array}\!\!\!\right\}}
\nc{\sbinlr}[2]{\Big\langle\!\!\begin{array}{c} {#1}\\ {#2} 
\end{array}\!\!\Big\rangle}
\nc{\bino}[2]{\left(\!\!\begin{array}{c} {#1}\\ {#2} \end{array}\!\!\right)}
\def\half {\mbox{$\textstyle \frac{1}{2}$}}

\definecolor{lightblue}{rgb}{.61,.61,1}
\definecolor{midblue}{rgb}{.7,.7,1}
\definecolor{lightlightblue}{rgb}{.85,.85,1}
\definecolor{lightestblue}{rgb}{.96,.96,1}
\definecolor{lightpurple}{rgb}{1,.65,1}
\nc{\ch}{{\rm ch}}
\nc{\R}{{\cal R}}
\nc{\dkk}{\delta_{j,\{k,k'\}}^{(2)}}
\nc{\drr}{\delta_{j,\{r,r'\}}^{(2)}}
\nc{\ddkk}{\delta_{j,\{k,k'\}}^{(4)}}
\nc{\dddkk}{\delta_{j,\{k,k'\}}^{(8)}}
\nc{\dnn}{\delta_{j,\{n,n'\}}^{(2)}}
\nc{\ddnn}{\delta_{j,\{n,n'\}}^{(4)}}
\nc{\dddnn}{\delta_{j,\{n,n'\}}^{(8)}}

\definecolor{pink}{rgb}{1,.65,.65}

\begin{document}

\topmargin -5mm
\oddsidemargin 5mm

\setcounter{page}{1}

\vspace{8mm}
\begin{center}
{\Large {\bf Grothendieck ring and Verlinde-like formula}}\\[.25cm]
{\Large {\bf for the ${\cal W}$-extended logarithmic minimal model ${\cal WLM}(1,p)$}}

\vspace{8mm}
 {\Large Paul A. Pearce\footnote{Email: {\tt P.Pearce@ms.unimelb.edu.au}}},
 \ \  {\Large J{\o}rgen Rasmussen\footnote{Email: {\tt J.Rasmussen@ms.unimelb.edu.au}}}
\\[.3cm]
 {\em Department of Mathematics and Statistics, University of Melbourne}\\
 {\em Parkville, Victoria 3010, Australia}
\\[.4cm]
 {\Large Philippe Ruelle\footnote{Email: {\tt philippe.ruelle@uclouvain.be}}}
\\[.3cm]
 {\em Institut de Physique Th\'eorique, Universit\'e catholique de Louvain}\\
 {\em 1348 Louvain-la-Neuve, Belgium}\\[.4cm]

\end{center}

\vspace{8mm}
\centerline{{\bf{Abstract}}}
\vskip.4cm
\noindent
We consider the Grothendieck ring of the fusion algebra of the ${\cal W}$-extended 
logarithmic minimal model ${\cal WLM}(1,p)$. Informally, this is the fusion ring 
of ${\cal W}$-irreducible characters so it is blind to the Jordan block structures associated 
with reducible yet indecomposable representations. 
As in the rational models, the Grothendieck ring is described by a simple graph 
fusion algebra.
The $2p$-dimensional matrices of the regular representation are mutually commuting 
but not diagonalizable. 
They are brought simultaneously to Jordan form by the modular data coming from the full 
$(3p-1)$-dimensional S-matrix which includes transformations of the $p-1$ pseudo-characters.  
The spectral decomposition yields a Verlinde-like formula that is manifestly independent of 
the modular parameter $\tau$ but is, in fact, equivalent to the Verlinde-like formula 
recently proposed by Gaberdiel and Runkel involving a $\tau$-dependent S-matrix.
\renewcommand{\thefootnote}{\arabic{footnote}}
\setcounter{footnote}{0}

\section{Introduction}

Fusion algebras carry fundamental and physically important information about the 
structure of conformal field theories (CFTs). In their most succinct form, fusion rules are 
encoded in graph fusion algebras~\cite{BPPZ}. 
For rational CFTs, the celebrated Verlinde formula~\cite{Ver88} states that the 
{\em modular S-matrix diagonalizes the fusion rules}. 
More precisely, the modular S-matrix is the similarity matrix which simultaneously 
brings the graph fusion matrices (in the regular 
representation) into diagonal form. Recall that, in the rational setting, the modular 
S-matrix specifies 
the linear transformations of the finite number of irreducible characters under the 
transformation $\tau \to -1/\tau$.
The properties of modular invariance~\cite{Cardy1} on the torus and modular 
covariance~\cite{Cardy2} in the presence of a boundary subsequently played a major 
role in the understanding of rational CFTs. In particular, it led to the Cardy formula for the 
cylinder partition functions in terms of the modular S-matrix thus establishing the boundary 
operator content of rational CFTs. 

The Verlinde and Cardy formulas~\cite{Car89,BPPZ} play a central role in the general 
formulation of rational CFTs. 
However, the CFTs describing systems such as polymers and critical percolation, 
are not rational CFTs --- they are logarithmic CFTs.
With the recent upsurge of interest in logarithmic CFTs~\cite{Flo03,Gab03}, 
a natural question concerns the generalized form of the Verlinde and Cardy formulas 
in the context of logarithmic CFTs. 
This poses a number of serious challenges: (i) the 
appearance of indecomposable representations whose characters are linear 
combinations of irreducible characters, (ii) the fact that the irreducible 
characters transform under $\tau \to -1/\tau$ through a $\tau$-dependent S-matrix, 
and (iii) the non-diagonalizability of fusion matrices. 
A more detailed account of these difficulties can be found in~\cite{FHST}.
The Verlinde and Cardy formulas are concerned with spectra and are blind to the Jordan 
block structures associated with indecomposability which is a characteristic of 
logarithmic theories. 
For this reason, it suffices to work with the Grothendieck ring rather than the full 
fusion ring of representations. 
In the cases of interest here,
the Grothendieck ring is the quotient fusion algebra obtained by elevating the various
character identities to equivalence relations between the corresponding representations.
Informally, it is the {\em fusion ring of irreducible characters}.
While this effectively solves problem (i), the other two problems need to be resolved. 

In this paper, we consider logarithmic minimal models~\cite{PRZ0607,RP0707}
in the ${\cal W}$-extended picture~\cite{PRR0803,Ras0805}, namely, the models 
${\cal WLM}(1,p)$ for integer $p>1$.   
The case $p=2$ is symplectic fermions~\cite{Kausch}. 
For these models, we present a derivation of a Verlinde-like formula close in spirit to the 
familiar treatment of 
the Verlinde formula in the rational setting which is based on spectral decompositions.  
The central charge of the ${\cal WLM}(1,p)$ model is $c= 13 - 6(p+\frac{1}{p})$ and the 
chiral vertex algebra ${\cal W}_p$ is generated by three fields of dimension $2p-1$, with 
$2p$ irreducible representations. 
The associated $2p$-dimensional Grothendieck fusion matrices are, in general, 
not diagonalizable and typically contain Jordan blocks of rank 2. 
We show that a single similarity matrix $Q$ simultaneously brings these matrices to 
Jordan form, albeit not necessarily strict Jordan {\em canonical} form. 
The matrix $Q$ is simply related to the modular matrix S. 
Specifically, the independent columns of the S-matrix provide a system of eigenvectors
and generalized eigenvectors which form the columns of $Q$.
Moreover, the eigenvalues of the Grothendieck fusion matrices are simple fractions 
in the entries of the S-matrix. It is in this sense, that our Verlinde-like formula reduces to a 
spectral decomposition of the Grothendieck fusion matrices, for which the modular 
S-matrix provides the eigendata. We stress, though, that the matrix $Q$ and 
its realization in terms of modular data are far from unique.

A complicating factor when studying ${\cal WLM}(1,p)$ is that, strictly speaking, 
the set of irreducible characters does not close under the modular transformation $\tau \to -1/\tau$. 
One can work with a $2p$-dimensional but $\tau$-dependent 
S-matrix (it is linear in $\tau$), 
or one can alternatively enlarge the space of characters by introducing $p-1$ so-called 
pseudo-characters~\cite{Flo9509,Flo9605,FG0509,FGST0606b} thereby obtaining a proper 
representation of the modular group. In either formulation, the S-matrices contain 
equivalent modular data. We work with the proper S-matrix but 
present our final result for the Verlinde-like formula in both formulations and so establish 
the equivalence of our Verlinde-like formula to the formula obtained recently by Gaberdiel 
and Runkel~\cite{GR0707}. Although their form of the Verlinde-like formula involves 
$\tau$, it was checked numerically to 
be independent of $\tau$ and to correctly reproduce the Grothendieck ring structure constants. 
We confirm this analytically and show that our Verlinde-like formula
reproduces the $\tau$-independent part of their formula  
thereby completing the proof of their formula.

Other approaches to a Verlinde-like formula for the ${\cal WLM}(1,p)$ models have also been 
proposed. In \cite{FHST}, the problem (ii) is circumvented 
by introducing automorphy factors in order to obtain a true linear representation 
of the modular group. The matrix representing the transformation $\tau \to -1/\tau$ is 
shown to block-diagonalize the Grothendieck fusion matrices, the block-diagonal form 
being expressible in terms of the entries of this same matrix. 
Albeit different from our formula, the resulting formula correctly reproduces the fusion 
coefficients (in the Grothendieck ring). As in \cite{GR0707}, the modular matrix is, 
in principle, $\tau$-dependent but the working requires the specific choice $\tau=i$. 
This block-diagonalization similarly misses the Jordan form which we believe is key to a 
proper interpretation of a Verlinde-like formula in a logarithmic setting.
Specifically, the modular S-matrix enters by providing the eigenvectors and generalized
eigenvectors that form the columns of the similarity matrix bringing the fusion matrices to 
Jordan form. The article \cite{FK07} generalizes the approach of \cite{FHST} to include 
the indecomposable representations and proposes an equivalent limit-Verlinde-like formula.
In a separate development~\cite{GT0711}, a Verlinde-like formula based on the {\em union}
of irreducible characters and pseudo-characters has been derived.

Finally, in a companion paper~\cite{Ras0906}, our present study of the Grothendieck
ring of ${\cal WLM}(1,p)$ is extended to the fusion algebra itself. The analysis is
very similar though technically more involved due to the presence of Jordan blocks
of rank 3 in the spectral decompositions of the fusion matrices. 
This also affects the form of the resulting Verlinde-like formulas.

\section{The logarithmic minimal model ${\cal WLM}(1,p)$}

A ${\cal W}$-extended logarithmic minimal model ${\cal WLM}(1,p)$ is 
defined~\cite{PRR0803,Ras0805} for every positive integer $p>1$.
It has central charge
\be
 c\ =\ 13-6\big(p+\frac{1}{p}\big)
\ee
while the conformal weights are given by
\be
 \D_{r,s}\ =\ \frac{(rp-s)^{2}-(p-1)^2}{4p},\hspace{1.2cm} r,s\in\mathbb{N}
\label{D}
\ee
There are $2p$ irreducible representations
\be
 \{\ketw{r,s};\ r\in\mathbb{Z}_{1,2},s\in\mathbb{Z}_{1,p}\},\qquad\quad
  \D[\ketw{r,s}]\ =\ \D_{r,s}
\ee
and $2p-2$ indecomposable rank-2 representations
\be
 \{\ketw{\R_{r}^{b}};\ r\in\mathbb{Z}_{1,2},b\in\mathbb{Z}_{1,p-1}\}
\ee
where we have introduced the notation $\mathbb{Z}_{n,m}=\mathbb{Z}\cap[n,m]$.
The irreducible characters are given by
\be
  \chih_{r,s}(q)\ =\ \chit[\ketw{r,s}](q)\ =\ 
     \frac{1}{\eta(q)}\sum_{j\in\mathbb{Z}}(2j+r)q^{p(j+\frac{r p-s}{2p})^2}
\label{Wirrchar}
\ee
where $\eta(q)$ is the Dedekind eta function
\be
 \eta(q)\ =\ q^{1/24}\prod_{m=1}^\infty(1-q^m)
\label{eta}
\ee
The characters of the indecomposable rank-2 representations are given by
\be
 \chit[\ketw{\R_{1}^{b}}](q)\ =\ \chit[\ketw{\R_{2}^{p-b}}](q)\ =\ 
  2\chih_{2,b}(q)+2\chih_{1,p-b}(q)
\label{chiR}
\ee
There are $p+1$ so-called projective characters. These are the characters $\chih_{r,p}(q)$ and
(\ref{chiR}) of the projective representations 
\be
 \{\ketw{r,p},\ \ketw{\R_r^b};\ r\in\mathbb{Z}_{1,2},b\in\mathbb{Z}_{1,p-1}\} 
\ee

\subsection{Fusion algebra}

The fundamental fusion algebra
\be
 \mathrm{Fund}[{\cal WLM}(1,p)]\ =\ 
   \big\langle\ketw{r,s},\ketw{\R_r^b};\  r\in\mathbb{Z}_{1,2},
     s\in\mathbb{Z}_{1,p}, b\in\mathbb{Z}_{1,p-1}\big\rangle
\label{fusalg}
\ee
of ${\cal WLM}(1,p)$ is commutative and associative. The underlying fusion rules
read~\cite{GK9606,EF0604,GR0707,PRR0803} 
\bea
 \ketw{r,s}\otimes\ketw{r',s'}&=&
  \!\bigoplus_{j=|s-s'|+1,\ \!\mathrm{by}\ \!2}^{p-|p-s-s'|-1}
  \!\!\!\ketw{r\cdot r',j}
    \oplus\!\!\bigoplus_{\beta=\eps(s+s'-p-1),\ \!\mathrm{by}\ \!2}^{s+s'-p-1}
    \!\!\!\ketw{\R_{r\cdot r'}^{\beta}}   
   \nn
 \ketw{r,s}\otimes\ketw{\R_{r'}^{b}}&=&
    \!\bigoplus_{\beta=|s-b|+1,\ \!\mathrm{by}\ \!2}^{p-|p-s-b|-1}
      \!\!\!\!\ketw{\R_{r\cdot r'}^{\beta}}
     \oplus\!\!\bigoplus_{\beta=\eps(s-b-1),\ \!\mathrm{by}\ \!2}^{s-b-1}
        \!\!\!\!2\ketw{\R_{r\cdot r'}^{\beta}}\oplus
     \!\!\!\bigoplus_{\beta=\eps(s+b-p-1),\ \!\mathrm{by}\ \!2}^{s+b-p-1}
        \!\!\!\!2\ketw{\R_{2\cdot r\cdot r'}^{\beta}}  
    \nn
 \ketw{\R_{r}^{b}}\otimes\ketw{\R_{r'}^{b'}}
  &=&\bigoplus_{\beta=\eps(p-b-b'-1),\ \!\mathrm{by}\ \!2}^{p-|b-b'|-1}
     \!\!\!\!2\ketw{\R_{r\cdot r'}^{\beta}}
     \oplus\!\!\bigoplus_{\beta=\eps(p-b-b'-1),\ \!\mathrm{by}\ \!2}^{|p-b-b'|-1}
     \!\!\!2\ketw{\R_{r\cdot r'}^{\beta}}\nn
   &&\qquad\oplus\!\bigoplus_{\beta=\eps(b+b'-1),\ \!\mathrm{by}\ \!2}^{p-|p-b-b'|-1}
     \!\!\!2\ketw{\R_{2\cdot r\cdot r'}^{\beta}}
     \oplus\!\bigoplus_{\beta=\eps(b+b'-1),\ \!\mathrm{by}\ \!2}^{|b-b'|-1}
     \!\!\!2\ketw{\R_{2\cdot r\cdot r'}^{\beta}}  
\label{fus}
\eea
where we have introduced $\ketw{\R_{r}^0}\equiv\ketw{r,p}$ and
\be
 \eps(n)\ =\ \frac{1-(-1)^n}{2},\qquad\quad
   n\cdot m\ =\ \frac{3-(-1)^{n+m}}{2},\qquad\quad n,m\in\mathbb{Z}
\label{eps}
\ee

\subsection{Fusion matrices and polynomial fusion ring}

We let
\be
 \big\{N_{\ketw{r,s}},N_{\ketw{\R_r^b}};\  r\in\mathbb{Z}_{1,2},
     s\in\mathbb{Z}_{1,p}, b\in\mathbb{Z}_{1,p-1}\big\}
\label{NN}
\ee 
denote the set of fusion matrices realizing
the fusion algebra (\ref{fus}) of ${\cal WLM}(1,p)$. These are all $(4p-2)$-dimensional
square matrices. Special notation is introduced for the two fundamental
fusion matrices
\be
 X\ =\ N_{\ketw{2,1}},\qquad\quad Y\ =\ N_{\ketw{1,2}}
\label{XY}
\ee
{}From~\cite{Ras0812}, we have
\be
 N_{\ketw{r,s}}\ =\ X^{r-1}U_{s-1}\big(\frac{Y}{2}\big),\qquad
 N_{\ketw{\R_r^b}}\ =\ 2X^{r-1}T_b\big(\frac{Y}{2}\big)U_{p-1}\big(\frac{Y}{2}\big)
\label{N}
\ee
where $T_n$ and $U_n$ are Chebyshev polynomials of the first and second kind, respectively. 
It follows from the refinement in~\cite{Ras0906} of the discussion of the associated 
quotient polynomial fusion ring in~\cite{Ras0812} that 
\be
 \mathrm{Fund}[{\cal WLM}(1,p)]\ \simeq\ 
   \mathbb{C}[X,Y]/\big(X^2-1,\tilde{P}_{1,p}(X,Y)\big)
\label{iso}
\ee
where
\be
  \tilde{P}_{1,p}(X,Y)\ =\ \Big(X-T_{p}\big(\frac{Y}{2}\big)\Big)U_{p-1}\big(\frac{Y}{2}\big)
\label{PP}
\ee

\subsection{Modular S-matrix}

The set of irreducible characters (\ref{Wirrchar}) does not close under modular
transformations. Instead, 
a representation of the modular group is obtained~\cite{FGST0606b} by enlarging
this set with the $p-1$ {\em pseudo-characters}
\be
 \chih_{0,b}(q)\ =\ i\tau\big(b\chih_{1,p-b}(q)-(p-b)\chih_{2,b}(q)\big)
\label{pseudo}
\ee
where the modular parameter is
\be
 q\ =\ e^{2\pi i\tau}
\label{qtau}
\ee
Writing the associated modular S-matrix in block form with respect to the distinction
between proper characters $\chih_{r,s}(q)$ and pseudo-characters $\chih_{0,b}(q)$, 
the matrix is
\be
 S\ =\ \begin{pmatrix}S_{r,s}^{r'\!,s'}&S_{r,s}^{0,b'}\\[4pt]
  S_{0,b}^{r'\!,s'}&S_{0,b}^{0,b'}\end{pmatrix}\ =\
  \begin{pmatrix}\frac{(2-\delta_{s',p})(-1)^{r s'+ r' s+ r r' p}s\cos\frac{ss'\pi}{p}}{p\sqrt{2p}}&
  \frac{2(-1)^{r b'}\sin\frac{sb'\pi}{p}}{p\sqrt{2p}}\\[6pt]
  \frac{2(-1)^{r' b}(p-s')\sin\frac{bs'\pi}{p}}{\sqrt{2p}}&0\end{pmatrix}
\label{S}
\ee
This matrix is not symmetric and not unitary but satisfies $S^2=I$. We note that
\be
 S_{r,s}^{1,p-b}\ =\ S_{r,s}^{2,b}
\ee
implying that, under the modular transformation $\tau\to\frac{-1}{\tau}$, the $2p$ irreducible
characters transform into linear combinations of the $p+1$ projective characters (with expansion 
coefficients $S_{r,s}^{r',p}$ and $\half S_{r,s}^{2,b}$) and the $p-1$ 
pseudo-characters (with expansion coefficients $S_{r,s}^{0,b}$), only.
We also note that, formally,
\be
 S_{r,s}^{2,b}\ =\ \frac{\pa}{\pa\theta_b}S_{r,s}^{0,b},\qquad \theta_b\ =\ \frac{b\pi}{p}
\label{deriv}
\ee

Alternatively, one can introduce the $2p$-dimensional, $\tau$-dependent (and thus {\em improper})
S-matrix 
\be
 \Sc-i\tau\Sct
 \label{Sc}
\ee
(here written in calligraphic to distinguish it from the proper S-matrix in (\ref{S}))
obtained by expanding the pseudo-characters in terms of the irreducible characters.
Its entries thus read
\be
 \Sc_{r,s}^{r',s'}\ =\ S_{r,s}^{r',s'},\qquad
 \Sct_{r,s}^{r',s'}\ =\ \frac{2(-1)^{r s'+r's+rr' p}(p-s')
   \sin\frac{ss'\pi}{p}}{p\sqrt{2p}}
\label{S0S1}
\ee
from which it follows that
\be
 (p-b)\Sct_{r,s}^{1,p-b}\ =\ -b\Sct_{r,s}^{2,b} 
\ee
and
\be
 \Sct_{r,s}^{1,b'}\ =\ -(p-b')S_{r,s}^{0,p-b'},\qquad
 \Sct_{r,s}^{2,b'}\ =\ (p-b')S_{r,s}^{0,b'},\qquad
 \Sct_{r,s}^{r',p}\ =\ 0
\ee
It is easily seen that
an expression can be written in terms of the proper S-matrix $S$ if and only if it can be 
written in terms of the improper S-matrix $\Sc-i\tau\Sct$.

\section{Grothendieck ring of ${\cal WLM}(1,p)$}

The Grothendieck ring of ${\cal WLM}(1,p)$ is obtained by elevating the
character identities of ${\cal WLM}(1,p)$ to relations between the corresponding generators
of the fusion algebra $\mathrm{Fund}[{\cal WLM}(1,p)]$ of ${\cal WLM}(1,p)$. From (\ref{chiR}),
we thus impose the equivalence relations
\be
  \ketw{\R_{1}^{b}}\ \sim\ \ketw{\R_{2}^{p-b}}\ \sim\ 
  2\ketw{2,b}\oplus2\ketw{1,p-b},\qquad\quad b\in\mathbb{Z}_{1,p-1}
\label{Gequiv}
\ee
Following (\ref{fus}), it is straightforwardly verified that 
\be
 \mathrm{Grot}[{\cal WLM}(1,p)]:=\
   \big\langle G_{r,s};\ r\in\mathbb{Z}_{1,2},
     s\in\mathbb{Z}_{1,p-1}\big\rangle_{1,p}\nn 
  \ \simeq\
   \mathrm{Fund}[{\cal WLM}(1,p)]\big/\!\sim
\ee
where the equivalence relation $\sim$ is defined in (\ref{Gequiv}) and where the rules with 
respect to the Grothendieck multiplication $\ast$ are
\be
  G_{r,s}\ast G_{r',s'}\ =\ 
  \!\sum_{j=|s-s'|+1,\ \!\mathrm{by}\ \!2}^{p-|p-s-s'|-1}
  \!\!\! G_{r\cdot r',j}
    +\!\!\sum_{\beta=\eps(s+s'-p-1),\ \!\mathrm{by}\ \!2}^{s+s'-p-1}
    \!\!\!(2-\delta_{\beta,0})\big(G_{r\cdot r',p-\beta}+G_{2\cdot r\cdot r',\beta}\big)   
\label{Gmult}
\ee
Here we are using $G_{r,0}\equiv0$.
Since
\be
 N_{\ketw{\R_r^b}}\ =\ 2X^{r-1}\Big(T_p\big(\frac{Y}{2}\big)U_{b-1}\big(\frac{Y}{2}\big)
   +U_{p-b-1}\big(\frac{Y}{2}\big)\Big)
\ee
and
\be
  N_{\ketw{2\cdot r,b}}+N_{\ketw{r,p-b}}\ \equiv\ 
    X^{r-1}\Big(XU_{b-1}\big(\frac{Y}{2}\big)+U_{p-b-1}\big(\frac{Y}{2}\big)\Big)\qquad
   (\mathrm{mod}\ X^2-1)
\ee
it follows (from setting $r=b=1$, in particular) that
\be
 \mathrm{Grot}[{\cal WLM}(1,p)]\ \simeq\  \mathbb{C}[X,Y]/\big(X^2-1,X-T_p\big(\frac{Y}{2}\big)\big)
  \ \simeq\ \mathbb{C}[Y]/\big((Y^2-4)U_{p-1}^2\big(\frac{Y}{2}\big)\big)
\label{Grot}
\ee
With reference to (\ref{iso}), it is noted that $X-T_p\big(\frac{Y}{2}\big)$ is a divisor of 
$\tilde{P}_{1,p}(X,Y)$. The first isomorphism in (\ref{Grot}) is given by
\be
 G_{2,1}\ \leftrightarrow\ X,\qquad G_{1,2}\ \leftrightarrow\ Y,\qquad
  G_{r,s}\ \leftrightarrow\ X^{r-1}U_{s-1}\big(\frac{Y}{2}\big)
\ee
while the second isomorphism is due to
\be
 X^2-1\ \equiv\ T_p^2\big(\frac{Y}{2}\big)-1\ =\ \frac{1}{4}(Y^2-4)U_{p-1}^2\big(\frac{Y}{2}\big)
 \qquad (\mathrm{mod}\ X-T_p\big(\frac{Y}{2}\big))
\ee
and allows us to write
\be
 G_{r,s}\ \leftrightarrow\ T_p^{r-1}\big(\frac{Y}{2}\big)U_{s-1}\big(\frac{Y}{2}\big)
\label{Gks}
\ee

\subsection{Graph fusion algebra of the Grothendieck ring}
\label{SecGraphG}

The Grothendieck ring of ${\cal WLM}(1,p)$ is described by a graph fusion algebra
\be
 N_i N_j\ =\ \sum_{k=1}^{2p} N_{ij}{}^k N_k
\ee
with $N_1=N_{1,1}=I$ and $N_2=N_{1,2}=Y$ the adjacency matrix of the 
fundamental graph as shown in Figure~1. 
The fusion matrices $N_i$ are mutually commuting but in general not symmetric, not normal 
and not diagonalizable. 
Nevertheless, we will show that they can be simultaneously brought to Jordan form 
by a similarity transformation determined from the modular data.
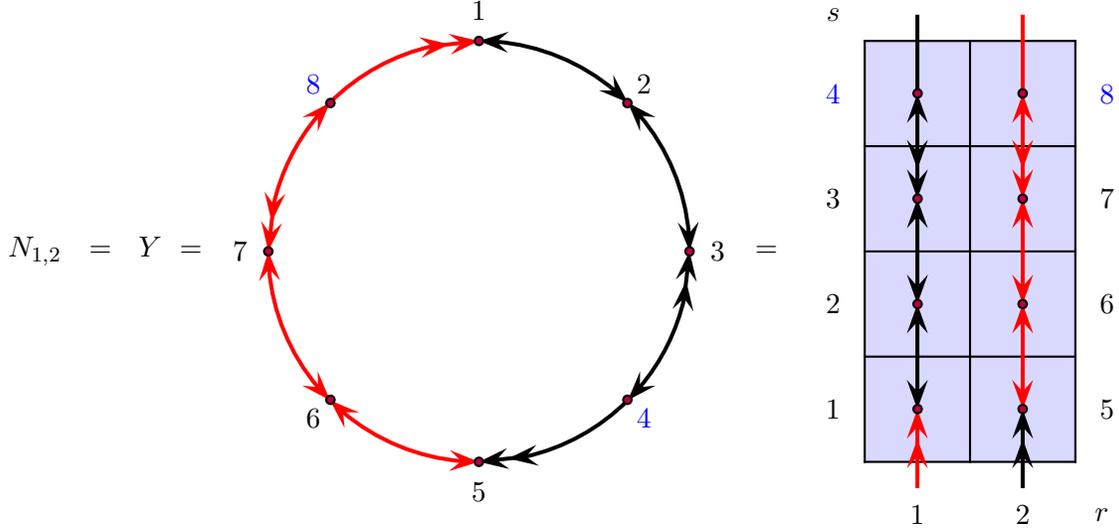
\begin{figure}[thbp]
\bea
\psset{unit=1.5cm}
N_{1,2}\ \ =\ \  Y \  =\qquad
\psset{unit=1.4cm}
\begin{pspicture}[shift=-1.95](4,4)
\psarc[linewidth=1.5pt,linecolor=black,arrowsize=8pt]{<->}(2,2){2}{0}{45}
\psarc[linewidth=1.5pt,linecolor=black,arrowsize=8pt]{<->}(2,2){2}{45}{90}
\psarc[linewidth=1.5pt,linecolor=red,arrowsize=8pt]{<<-}(2,2){2}{90}{135}
\psarc[linewidth=1.5pt,linecolor= red,arrowsize=8pt]{<->>}(2,2){2}{135}{180}
\psarc[linewidth=1.5pt,linecolor= red,arrowsize=8pt]{<->}(2,2){2}{180}{225}
\psarc[linewidth=1.5pt,linecolor= red,arrowsize=8pt]{<->}(2,2){2}{225}{270}
\psarc[linewidth=1.5pt,linecolor=black,arrowsize=8pt]{<<-}(2,2){2}{270}{315}
\psarc[linewidth=1.5pt,linecolor=black,arrowsize=8pt]{<->>}(2,2){2}{315}{360}
\pscircle[fillstyle=solid,fillcolor=purple](4,2){.05}
\pscircle[fillstyle=solid,fillcolor=purple](3.41,3.41){.05}
\pscircle[fillstyle=solid,fillcolor=purple](2,4){.05}
\pscircle[fillstyle=solid,fillcolor=purple](.59,3.41){.05}
\pscircle[fillstyle=solid,fillcolor=purple](0,2){.05}
\pscircle[fillstyle=solid,fillcolor=purple](.59,.59){.05}
\pscircle[fillstyle=solid,fillcolor=purple](2,0){.05}
\pscircle[fillstyle=solid,fillcolor=purple](3.41,.59){.05}
\rput[b](2,4.2){1}
\rput[bl](3.5,3.5){2}
\rput[l](4.2,2){3}
\rput[tl](3.5,.5){\color{blue}4}
\rput[t](2,-.2){5}
\rput[tr](.5,.5){6}
\rput[r](-.2,2){7}
\rput[br](.5,3.5){\color{blue}8}
\end{pspicture}
\qquad=
\begin{pspicture}[shift=-2.45](-.75,-.5)(2,4.25)
\psframe[linewidth=0pt,fillstyle=solid,fillcolor=lightlightblue](0,0)(2,4)
\psgrid[gridlabels=0pt,subgriddiv=1](0,0)(2,4)
\psline[linewidth=1.5pt,linecolor=red,arrowsize=8pt]{->>}(.5,-.25)(.5,.5)
\psline[linewidth=1.5pt,linecolor=black,arrowsize=8pt]{<->}(.5,.5)(.5,1.5)
\psline[linewidth=1.5pt,linecolor=black,arrowsize=8pt]{<->}(.5,1.5)(.5,2.5)
\psline[linewidth=1.5pt,linecolor=black,arrowsize=8pt]{<<->}(.5,2.5)(.5,3.5)
\psline[linewidth=1.5pt,linecolor=black,arrowsize=8pt]{-}(.5,3.5)(.5,4.25)
\psline[linewidth=1.5pt,linecolor=black,arrowsize=8pt]{->>}(1.5,-.25)(1.5,.5)
\psline[linewidth=1.5pt,linecolor=red,arrowsize=8pt]{<->}(1.5,.5)(1.5,1.5)
\psline[linewidth=1.5pt,linecolor= red,arrowsize=8pt]{<->}(1.5,1.5)(1.5,2.5)
\psline[linewidth=1.5pt,linecolor= red,arrowsize=8pt]{<<->}(1.5,2.5)(1.5,3.5)
\psline[linewidth=1.5pt,linecolor= red,arrowsize=8pt]{-}(1.5,3.5)(1.5,4.25)
\multiput(.5,.5)(0,1){4}{\pscircle[fillstyle=solid,fillcolor=purple](0,0){.05}}
\multiput(1.5,.5)(0,1){4}{\pscircle[fillstyle=solid,fillcolor=purple](0,0){.05}}
{\color{black}
\rput(.5,-.5){$1$}
\rput(1.5,-.5){$2$}
\rput(2.25,-.5){$r$}
\rput(-.3,.5){$1$}
\rput(-.3,1.5){$2$}
\rput(-.3,2.5){$3$}
\rput(2.3,.5){$5$}
\rput(2.3,1.5){$6$}
\rput(2.3,2.5){$7$}
\rput(-.3,4.25){$s$}
\color{blue}\rput(-.3,3.5){$4$}
\color{blue}\rput(2.3,3.5){$8$}}
\end{pspicture}\nonumber
\eea
\caption{The fundamental fusion graph $Y=N_{1,2}$ of the Grothendieck ring of 
${\cal WLM}(1,4)$ and the same graph superimposed on the Kac table. 
The labels are $1=(1,1)$, $2=(1,2)$, $3=(1,3)$, $4=(1,4)$, $5=(2,1)$, $6=(2,2)$, 
$7=(2,3)$ and $8=(2,4)$. 
The fundamental fusion graphs for larger values of $p$ are obtained by adding 
additional doubly directed bonds at the positions labelled 1 and 5.}
\end{figure}

The regular representation of the graph fusion algebra of the Grothendieck ring of 
${\cal WLM}(1,p)$ is specified in terms of $2p$-dimensional matrices. 
In the ordered basis 
\be
 G_{1,1},G_{2,1};\ldots;G_{1,s},G_{2,s};\ldots;G_{1,p},G_{2,p}
\label{order}
\ee
the {\em fundamental} Grothendieck matrices $N_{2,1}=X$ and $N_{1,2}=Y$ are given by
\be
 X\ =\ C_{2p},\qquad\quad Y\ =\ 
\renewcommand{\arraystretch}{1.2}
\left(\!\!
\begin{array}{cccccc|c}
  0&I&0&\cdots&&&0 \\
  I&0&I&\ddots&&& \\
  0&I&0&\ddots&&&\vdots \\
  \vdots&\ddots&\ddots&\ddots&&& \\
  &&&&0&I&0 \\
  0&\cdots&\cdots&0&I&0&I \\
\hline
  2C&0&\cdots&\cdots&0&2I&0
\end{array}
\!\!\right)
\label{G21G12}
\ee
where
\be
 C_{2p}\ =\ \mathrm{diag}(\underbrace{C,\ldots,C}_{p}),\qquad\quad
   C\ =\ \left(\!\!\begin{array}{cc} 0&1 \\ 1&0\end{array}\!\!\right)
\ee
is an involutory matrix, $C_{2p}^2=I$. 
The $2p$-dimensional matrix $Y$ is written here as a $p$-dimensional matrix, with 
$2\times2$ matrices $0=\left(\!\!\begin{array}{cc} 0&0 \\ 0&0\end{array}\!\!\right)$ 
and $I=\left(\!\!\begin{array}{cc} 1&0 \\ 0&1\end{array}\!\!\right)$ as entries, 
whose $p$'th row and column are emphasized to indicate their special status.
For $p=2$, the expression (\ref{G21G12}) for $Y$ is meant to reduce to
\bea
 Y\Big|_{p=2}\ =\ \left(\!\!\begin{array}{cc} 0&I \\ 2C+2I&0
    \end{array}\!\!\right)\ =\ \left(\!\!\begin{array}{cccc} 0&0&1&0 \\ 0&0&0&1\\ 2&2&0&0 
      \\ 2&2&0&0
    \end{array}\!\!\right)
\label{G12p2}
\eea

\subsection{Spectral decomposition}

The minimal and characteristic polynomials of $X$ are readily seen to be
\be
 X^2-I\ =\ (X-I)(X+I),\qquad\quad  \det(\la I-X)\ =\ (\la-1)^{p}(\la+1)^{p}
\label{charX}
\ee
The eigenvalues of $Y$ are
\be
 \beta_j\ =\ 2\al_j\ =\ 2\cos\theta_j,\qquad\quad \theta_j\ =\ \frac{j\pi}{p},\qquad j\in\mathbb{Z}_{0,p}
\label{betaY}
\ee
This follows from the observation, to be proven below, 
that the minimal and characteristic polynomials of $Y$ are
\bea
  (Y^2-4I)U_{p-1}^2\big(\frac{Y}{2}\big)
    &=&(Y-2I)(Y+2I)\prod_{b=1}^{p-1}\big(Y-\beta_bI\big)^2\nn
 \det(\la I-Y)&=&(\la-2)(\la+2)\prod_{b=1}^{p-1}\big(\la-\beta_b\big)^2
\label{charY}
\eea
where we have used that the Chebyshev polynomials of the second kind factorize as
\be
 U_{p-1}(x)\ =\ 2^{p-1}\prod_{b=1}^{p-1}\big(x-\al_b\big)
\label{Ua}
\ee
The expressions (\ref{charY}) 
imply that the Jordan canonical form $J_Y$ of $Y$ consists of $p-1$ rank-2 blocks
associated to the eigenvalues $\beta_b$, $b\in\mathbb{Z}_{1,p-1}$, 
and a rank-1 block associated to each of the eigenvalues $\beta_0=2$ and $\beta_p=-2$.

\section{Verlinde-like formulas}

First, we note that the eigenvalues (\ref{betaY}) can be written in terms of modular data as  
\be
 \beta_{(r-1)p}\ =\ \frac{S_{1,2}^{r,p}}{S_{1,1}^{r,p}},\qquad\quad
 \beta_b\ =\ \frac{S_{1,2}^{0,b}}{S_{1,1}^{0,b}}
\ee
Second, we wish to Jordan decompose the matrix $Y$ using a similarity matrix
whose entries are given in terms of the modular data.
For $r'\in\mathbb{Z}_{1,2}$ and $b\in\mathbb{Z}_{1,p-1}$, we thus define the 
$2p$-dimensional vectors $v_{(r'-1)p}$, $v_b$ and $w_b$ whose entries are given by
\be
  \left[v_{(r'-1)p}\right]_{r,s}\ =\ S_{r,s}^{r',p},
   \qquad\quad
 \left[v_b\right]_{r,s}\ =\ S_{r,s}^{0,b},
   \qquad\quad
 \left[w_b\right]_{r,s}\ =\ S_{r,s}^{2,b}
\ee
The vector $v_j$, $j\in\mathbb{Z}_{0,p}$, is the unique eigenvector of $Y$ 
associated to the eigenvalue $\beta_j$, while the doublet $v_b$, $w_b$ 
forms the generalized Jordan chain
\be 
 Y\big(v_b\ w_b\big)\ =\ \big(v_b\ w_b\big)\left(\!\!\begin{array}{cc} \beta_b&-2\sin\theta_b
   \\ 0&\beta_b  \end{array}\!\!\right) 
\ee
These assertions are straightforwardly proven using standard properties of the 
Chebyshev polynomials,
such as their recurrence relations, and immediately imply the results (\ref{charY}).
It follows that the similarity matrix
\be
  Q\ =\ \left(\!\!\begin{array}{c|c|cc|c|c} v_0&\ldots&
    v_{b}&w_{b}&\ldots&v_p  \end{array}\!\!\right)
\label{Qh}
\ee
whose entries are given by
\be
 Q\ =\ \left(\!\!\begin{array}{c|c|cc|c|c} S_{r,s}^{1,p}&\ldots& S_{r,s}^{0,b}
    &S_{r,s}^{2,b}&\ldots&S_{r,s}^{2,p}  \end{array}\!\!\right)
\label{QS}
\ee
converts the matrix realization of the Grothendieck generator $Y$ 
into the (in general, non-canonical) Jordan form
\be
Q^{-1} Y Q\ =\ \mathrm{diag}\big(\beta_0,\ldots,
   \left(\!\!\begin{array}{cc} \beta_b&-2\sin\theta_b
     \\ 0&\beta_b  \end{array}\!\!\right),\ldots,\beta_p\big)
\ee
The entries of the inverse of the similarity matrix $Q$ are given by
\be
 Q^{-1}\ =\ \left(\!\!\begin{array}{c}
  S_{1,p}^{r,s} \\ \hline \vdots \\ \hline
     S_{0,b}^{r,s} 
  \\ \frac{S_{1,p}^{1,p}}{S_{1,b}^{1,p}}S_{2,b}^{r,s}
  \\ \hline \vdots \\ \hline 
    S_{2,p}^{r,s} \end{array}\!\!\right)  
\label{Qm1}
\ee
The columns are labeled by $\ketw{r,s}$, while the rows are labeled by
$v_0$, the $p-1$ doublets $v_b$, $w_b$, and $v_p$.
As an aside, we have verified for small values of $p$ (analytically for $p=2,3,4,5,6$ 
and numerically up to $p=30$) that the determinant of $Q$ is 
\be
 \mathrm{det}(Q)\ =\ \frac{(-1)^{\lfloor\frac{p-2}{2}\rfloor}}{p^{p-1}}
\ee
Finally, the use of modular data obtained from transforming the pseudo-characters, as in
the expression (\ref{Qm1}) for $Q^{-1}$, can be avoided since
\be
 S_{0,b'}^{r,s}\ =\ 
    \frac{S_{1,p}^{1,p}(S_{1,p}^{1,p}-S_{1,s}^{1,p})}{(S_{1,1}^{1,p})^2}S_{r,s}^{0,b'},\qquad\quad
 S_{0,b'}^{r,b}\ =\ \frac{S_{1,p}^{1,p}S_{1,p-b}^{1,p}}{(S_{1,1}^{1,p})^2}S_{r,b}^{0,b'},
   \qquad S_{0,b'}^{r,p}\ =\ 0
\ee

The matrices realizing the other generators can be brought to a similar form since, 
referring to (\ref{Gks}), we have
\be
 N_{r,s}\ =\ f_{r,s}(Y)
\ee
Here, we have introduced the $2p$ functions
\be
 f_{r,s}(x)\ =\ T_p^{r-1}\big(\frac{x}{2}\big)U_{s-1}\big(\frac{x}{2}\big)
\ee
satisfying
\be
  f_{r,s}(\beta_0)\ =\ s,\qquad
  f_{r,s}(\beta_b)\ =\ \frac{(-1)^{(r-1)b}\sin s\theta_b}{\sin\theta_b},\qquad
  f_{r,s}(\beta_p)\ =\ s(-1)^{(r-1)p+s-1}
\ee
Using the following formula valid for any regular function $f$,
\be
 f \begin{pmatrix} \lambda & \mu \\ 0 & \lambda \end{pmatrix}\ =\ 
  \begin{pmatrix} f(\lambda) & \mu f'(\lambda) \\ 0 & f(\lambda) \end{pmatrix}
\ee
we easily obtain the Jordan form for the matrices $N_{r,s}$,
\be
 J_{r,s}\ =\ Q^{-1}N_{r,s}Q\ =\ \mathrm{diag}\big(f_{r,s}(\beta_0),\ldots,
   \left(\!\!\begin{array}{cc} f_{r,s}(\beta_b)&-2\sin\theta_bf_{r,s}'(\beta_b)
     \\ 0&f_{r,s}(\beta_b)  \end{array}\!\!\right),\ldots,f_{r,s}(\beta_p)\big)
\label{Jh}
\ee
We emphasize that this Jordan form can be expressed as a single function
acting on a Jordan {\em canonical} form
\be
 J_{r,s}\ =\ f_{r,s}\circ\phi\Big(\mathrm{diag}\big(\theta_0,\Jc_{\theta_1,2},\ldots
   \Jc_{\theta_b,2},\ldots,\Jc_{\theta_{p-1},2},\theta_p\big)\Big),\qquad\phi(\theta)\ =\ 2\cos\theta
\ee
where
\be
 \Jc_{\lambda,2}\ =\ \left(\!\!\begin{array}{cc}\lambda&1\\ 0&\lambda\end{array}\!\!\right)
\ee

Since the eigenvalues of the matrices $N_{r,s}$ also can be written in terms of the 
modular data
\be
 f_{r,s}(\beta_0) \ =\ \frac{S_{r,s}^{1,p}}{S_{1,1}^{1,p}},\qquad   
 f_{r,s}(\beta_b)\ =\ \frac{S_{r,s}^{0,b}}{S_{1,1}^{0,b}},\qquad 
 f_{r,s}(\beta_p)\ =\ \frac{S_{r,s}^{2,p}}{S_{1,1}^{2,p}}
\ee
it follows that the announced generalization of the Verlinde formula can be written
\be
 N_{r,s}\ =\ Q\:
  \mathrm{diag}\Big(\frac{S_{r,s}^{1,p}}{S_{1,1}^{1,p}},\ldots,
  \left(\!\!\begin{array}{cc}\frac{S_{r,s}^{0,b}}{S_{1,1}^{0,b}}
    &\frac{S_{r,s}^{2,b}}{S_{1,1}^{0,b}}-
   \frac{S_{1,1}^{2,b}S_{r,s}^{0,b}}{(S_{1,1}^{0,b})^2}\\ 0
   &\frac{S_{r,s}^{0,b}}{S_{1,1}^{0,b}} \end{array}\!\!\right),\ldots,
    \frac{S_{r,s}^{2,p}}{S_{1,1}^{2,p}}\Big)\,
 Q^{-1}
 \label{ver1}
\ee
In terms of the formal derivative (\ref{deriv}), this formula can be expressed compactly as
\be
 N_{r,s}\ =\ Q\:
  \mathrm{diag}\Big(\frac{S_{r,s}^{1,p}}{S_{1,1}^{1,p}},\ldots,
  \frac{S_{r,s}^{0,b}}{S_{1,1}^{0,b}}(\Jc_{\theta_b,2}),\ldots,
    \frac{S_{r,s}^{2,p}}{S_{1,1}^{2,p}}\Big)\,
 Q^{-1}
 \label{ver1a}
\ee

Every entry of the matrix $N_{r,s}$ in (\ref{ver1})
can be interpreted as a sum of three contributions; one obtained by
summing over the projective characters, one over the pseudo-characters, and one 
off-diagonal term with a sum over both projective and pseudo-characters. We thus have
\be
 \left[N_{r,s}\right]_{r',s'}^{r'',s''}\ =\ 
  \left[N_{r,s}^{\mathrm{proj}}\right]_{r',s'}^{r'',s''}
 +  \left[N_{r,s}^{\mathrm{pseudo}}\right]_{r',s'}^{r'',s''}
 +  \left[N_{r,s}^{\mathrm{off}}\right]_{r',s'}^{r'',s''}
\label{Nppp}
\ee
where we have introduced
\bea
 \left[N_{r,s}^{\mathrm{proj}}\right]_{r',s'}^{r'',s''}
  &=&S_{r',s'}^{1,p}F_{r,s}^{1,p}S_{1,p}^{r'',s''}
  +\sum_{b=1}^{p-1}S_{r',s'}^{2,b}F_{r,s}^{2,b}S_{2,b}^{r'',s''}
  +S_{r',s'}^{2,p}F_{r,s}^{2,p}S_{2,p}^{r'',s''}\nn
 \left[N_{r,s}^{\mathrm{pseudo}}\right]_{r',s'}^{r'',s''}
  &=& \sum_{b=1}^{p-1}S_{r',s'}^{0,b}F_{r,s}^{0,b}S_{0,b}^{r'',s''},\qquad\quad
 \left[N_{r,s}^{\mathrm{off}}\right]_{r',s'}^{r'',s''}
  \ =\ \sum_{b=1}^{p-1}S_{r',s'}^{0,b}F_{r,s}^{0,b;2,b}S_{2,b}^{r'',s''}
\eea
and
\bea
 &&
 F_{r,s}^{1,p}\ =\ \frac{S_{1,p}^{1,p}S_{r,s}^{1,p}}{(S_{1,1}^{1,p})^2}
 ,\qquad\quad
 F_{r,s}^{2,b}\ =\ \frac{S_{1,p}^{1,p}S_{r,s}^{0,b}}{S_{1,b}^{1,p}S_{1,1}^{0,b}}
 ,\qquad\quad
 F_{r,s}^{2,p}\ =\ \frac{S_{1,p}^{1,p}S_{r,s}^{2,p}}{S_{1,1}^{1,p}S_{1,1}^{2,p}}
 \nn
 &&
 F_{r,s}^{0,b}\ =\ \frac{S_{r,s}^{0,b}}{S_{1,1}^{0,b}},\qquad\qquad
 F_{r,s}^{0,b;2,b}\ =\ 
  \frac{S_{1,p}^{1,p}(S_{1,1}^{0,b}S_{r,s}^{2,b}-S_{1,1}^{2,b}S_{r,s}^{0,b})}{S_{1,b}^{1,p}
    (S_{1,1}^{0,b})^2}
\eea

The Verlinde-like formula (\ref{ver1}) can also be written in terms of the improper S-matrix (\ref{Sc}). 
A bit of rewriting thus yields
\bea
 \left[N_{r,s}\right]_{r',s'}^{r'',s''} &=& 
  \sum_{\nu=1}^{2}\Sc_{r',s'}^{\nu,p}\frac{\Sc_{r,s}^{\nu,p}}{\Sc_{1,1}^{\nu,p}}\Sc_{\nu,p}^{r'',s''}
   +\sum_{b=1}^{p-1}\sum_{\nu=1}^2\Big(\Sc_{r',s'}^{\nu,b} 
   \frac{\Sct_{r,s}^{\nu,b}}{\Sct_{1,1}^{\nu,b}}\Sc_{\nu,b}^{r'',s''} 
   + \Sct_{r',s'}^{\nu,b} \frac{\Sct_{r,s}^{\nu,b}}{\Sct_{1,1}^{\nu,b}}\Sct_{\nu,b}^{r'',s''} \nn
 && \hspace{5.6cm} + \:  \Sct_{r',s'}^{\nu,b}
  \Big[\frac{\Sc_{r,s}^{\nu,b} \Sct_{1,1}^{\nu,b} 
       -\Sc_{1,1}^{\nu,b} \Sct_{r,s}^{\nu,b}}{(\Sct_{1,1}^{\nu,b})^2}\Big]\Sc_{\nu,b}^{r'',s''}\Big)
\label{ver2}
\eea
A similar but superficially $\tau$-dependent expression was conjectured in~\cite{GR0707}.
We have managed to prove analytically that their formula is indeed $\tau$-independent and that
the manifestly $\tau$-independent part of their formula is equivalent to (\ref{ver2}).
In our notation, the proof of the $\tau$ independence amounts to showing that
\be
 \sum_{b=1}^{p-1}\sum_{\nu=1}^2\frac{\Sct_{r,s}^{\nu,b}\Sct_{r',s'}^{\nu,b}
  \Sc_{\nu,b}^{r'',s''}}{\Sct_{1,1}^{\nu,b}}\ =\ 0\ =\ 
  \sum_{b=1}^{p-1}\sum_{\nu=1}^2\frac{\Sct_{r,s}^{\nu,b}\Sct_{r',s'}^{\nu,b}
   \Sc_{1,1}^{\nu,b}\Sct_{\nu,b}^{r'',s''}}{(\Sct_{1,1}^{\nu,b})^2}
\ee

The Verlinde-like formula (\ref{ver2}) can be written in a slightly more compact form. 
{}From the explicit expressions (\ref{S0S1}) for the matrices $\Sc$ and $\Sct$, 
one can easily see that the inner ratio in the second term of (\ref{ver2}) has a 
well-defined limit when $t \to p$, equal to the inner ratio in the first term. 
So one accounts for the first term if one extends, for the second term, the summation 
from 1 to $p$. Moreover, the remaining two terms in (\ref{ver2}) do not change if we 
extend the summation to $p$, because the inner ratios are again well-defined in the 
limit $t \to p$, but are then multiplied by the entries $\Sct^{\nu,p}_{r',s'}=0$. 
We thus obtain a symmetrical expression, which contains a summation over all $2p$ 
indices, namely
\bea
 \left[N_{r,s}\right]_{r',s'}^{r'',s''} &=& \sum_{j=1}^{p} \sum_{\nu=1}^2\Big(
  \Sc_{r',s'}^{\nu,j}\frac{\Sct_{r,s}^{\nu,j}}{\Sct_{1,1}^{\nu,j}}\Sc_{\nu,j}^{r'',s''} 
  + \Sct_{r',s'}^{\nu,j}\frac{\Sct_{r,s}^{\nu,j}}{\Sct_{1,1}^{\nu,j}}\Sct_{\nu,j}^{r'',s''} \nn
 && \hspace{2cm} + \: \Sct_{r',s'}^{\nu,j}
   \Big[\frac{\Sc_{r,s}^{\nu,j} \Sct_{1,1}^{\nu,j} 
      -\Sc_{1,1}^{\nu,j} \Sct_{r,s}^{\nu,j}}{(\Sct_{1,1}^{\nu,j})^2} \Big]\Sc_{\nu,j}^{r'',s''}\Big)
\eea

\subsection{The case ${\cal WLM}(1,2)$}

The four-dimensional Grothendieck ring  
\be
 \mathrm{Grot}[{\cal WLM}(1,2)]
  \ \simeq\ \mathbb{C}[Y]/(Y^4-4Y^2)
\label{Grot2}
\ee
is generated by
\be
 G_{1,1}\ \leftrightarrow\ I,\qquad
 G_{1,2}\ \leftrightarrow\ Y,\qquad
 G_{2,1}\ \leftrightarrow\ \tfrac{1}{2}Y^2-I,\qquad
 G_{2,2}\ \leftrightarrow\ \tfrac{1}{2}Y^3-Y
\label{Gks2}
\ee
The multiplication rules are given in the Cayley tables in Figure~\ref{Cayley}.
\psset{unit=1cm}
\begin{figure}
$$
\renewcommand{\arraystretch}{1.5}
\begin{array}{c||cccc}
\ast&G_{1,1}&G_{2,1}&G_{1,2}&G_{2,2}\\[4pt]
\hline \hline
\rule{0pt}{14pt}
 G_{1,1}&G_{1,1}&G_{2,1}&G_{1,2}&G_{2,2}
    \\[4pt]
 G_{2,1}&G_{2,1}&G_{1,1}&G_{2,2}&G_{1,2}
    \\[4pt]
 G_{1,2}&G_{1,2}&G_{2,2}&2G_{1,1}+2G_{2,1}&2G_{1,1}+2G_{2,1}
    \\[4pt]
 G_{2,2}&G_{2,2}&G_{1,2}&2G_{1,1}+2G_{2,1}&2G_{1,1}+2G_{2,1}
\end{array}
\hspace{1.3cm}
\begin{array}{c||cccc}
\ast&0&1&-\tfrac{1}{8}&\tfrac{3}{8}\\[4pt]
\hline \hline
\rule{0pt}{14pt}
 0&0&1&-\tfrac{1}{8}&\tfrac{3}{8}
    \\[4pt]
 1&1&0&\tfrac{3}{8}&-\tfrac{1}{8}
    \\[4pt]
 -\tfrac{1}{8}&-\tfrac{1}{8}&\tfrac{3}{8}&2(0)+2(1)&2(0)+2(1)
    \\[4pt]
 \tfrac{3}{8}&\tfrac{3}{8}&-\tfrac{1}{8}&2(0)+2(1)&2(0)+2(1)
\end{array}
$$
\caption{Cayley tables of the multiplication rules for $\mathrm{Grot}[{\cal WLM}(1,2)]$. 
In the second table, the generators $G_{r,s}$ are represented by the corresponding conformal
weights.}
\label{Cayley}
\end{figure}
The fundamental fusion matrix $Y$ is given in (\ref{G12p2}).

There is only one pseudo-character, $\chih_{0,1}(q)=i\tau(\chih_{1,1}(q)-\chih_{2,1}(q))$,
and in the ordered basis 
$\{\chih_{1,1}(q),\chih_{2,1}(q),\chih_{1,2}(q),\chih_{2,2}(q),\chih_{0,1}(q)\}$,
the proper S-matrix is given by 
\be
 S\ =\ \begin{pmatrix}
   0&0&\tfrac{1}{4}&-\tfrac{1}{4}&-\tfrac{1}{2}\\[4pt]
   0&0&\tfrac{1}{4}&-\tfrac{1}{4}&\tfrac{1}{2}\\[4pt]
   1&1&\tfrac{1}{2}&\tfrac{1}{2}&0\\[4pt]
   -1&-1&\tfrac{1}{2}&\tfrac{1}{2}&0\\[4pt]
   -1&1&0&0&0
\end{pmatrix}
\label{Sp2}
\ee
The similarity matrix $Q$ (\ref{QS}) and its inverse $Q^{-1}$ (\ref{Qm1}) are given by
\be
 Q\ =\ \begin{pmatrix}
   \tfrac{1}{4}&-\tfrac{1}{2}&0&-\tfrac{1}{4}\\[4pt]
   \tfrac{1}{4}&\tfrac{1}{2}&0&-\tfrac{1}{4}\\[4pt]
   \tfrac{1}{2}&0&1&\tfrac{1}{2}\\[4pt]
   \tfrac{1}{2}&0&-1&\tfrac{1}{2}
\end{pmatrix},\qquad\quad
 Q^{-1}\ =\ \begin{pmatrix}
   1&1&\tfrac{1}{2}&\tfrac{1}{2}\\[4pt]
   -1&1&0&0\\[4pt]
   0&0&\tfrac{1}{2}&-\tfrac{1}{2}\\[4pt]
   -1&-1&\tfrac{1}{2}&\tfrac{1}{2}
\end{pmatrix}
\ee
They convert the four Grothendieck matrices $N_{r,s}$, $r,s\in\mathbb{Z}_{1,2}$,
simultaneously into the Jordan forms
\bea
 Q^{-1}N_{1,1}Q\ =\ \mathrm{diag}(1,1,1,1),&\qquad&
 Q^{-1}N_{1,2}Q\ =\ \mathrm{diag}\big(2,\begin{pmatrix} 0&-2\\ 0&0 \end{pmatrix},-2\big)\nn
 Q^{-1}N_{2,1}Q\ =\ \mathrm{diag}(1,-1,-1,1),&\qquad&
 Q^{-1}N_{2,2}Q\ =\ \mathrm{diag}\big(2,\begin{pmatrix} 0&2\\ 0&0 \end{pmatrix},-2\big)
\label{QNQp2}
\eea
It is noted that these are not all in Jordan {\em canonical} form.
The Verlinde-like formula (\ref{ver1}) follows by inverting the decompositions (\ref{QNQp2}).

\vskip.5cm
\section*{Acknowledgments}
\vskip.1cm
\noindent
This work is supported by the Australian Research Council and by the 
Belgian Interuniversity Attraction Poles Program P6/02, through the network 
NOSY (Nonlinear systems, stochastic processes and statistical mechanics). 
PR is a Research Associate of the Belgian National Fund for Scientific Research (FNRS).
PAP thanks the Asia Pacific Center for Theoretical Physics for hospitality during the early
stages of this work.


\end{document}